# On the Hamiltonian whose spectrum coincides with the set of primes

S. K. Sekatskii[1]

Laboratoire de Physique de la Matière Vivante, IPMC, BSP, Ecole Polytechnique Fédérale de Lausanne, CH-1015 Lausanne, Switzerland.

Institute of Spectroscopy Russian Academy of Sciences, Troitsk Moscow region, 142190 Russia.

**Abstract**

The problem of construction of a simple one – dimensional Hamiltonian $H = -\frac{d^2}{dx^2} + V(x)$ whose spectrum coincides with the set of primes is considered. We note that quasiclassically a Hamiltonian whose spectrum has the same counting function as that of the primes in the leading order (i. e. integral logarithm $\pi(N) \cong li(N)$) can be constructed with the function $V(x)$ whose inverse is asymptotically given by $x \cong li(V^{1/2})$. Hamiltonians, whose spectra coincide with the first 1 - 1000 primes are constructed numerically and their fractal structure is revealed. The possibility of using such a "prime – generating" Hamiltonian for investigation of certain number-theoretical problems is discussed.

---

[1] E-mail: Serguei.Sekatski@epfl.ch

## 1. Introduction

It is well known, that the Hamiltonian $H = -\frac{1}{2}\frac{d^2}{dx^2} + \frac{1}{2}x^2 + \frac{1}{2}$, which quantum – mechanically describes a one dimensional harmonic oscillator, has a spectrum coinciding with the set of integers. In other words, Schrödinger equation $(-\frac{1}{2}\frac{d^2}{dx^2} + \frac{1}{2}x^2 - \frac{1}{2})\psi = \lambda\psi$, or its equivalent form $(-\frac{d^2}{dx^2} + \frac{1}{4}x^2 - \frac{1}{2})\psi = \lambda\psi$, admits solutions from $L^2(-\infty, +\infty)$ iff $\lambda = 0, 1, 2, 3...$ This remarkable property naturally poses a question: is it possible to use spectral properties of this and other specially constructed Hamiltonians to *study and analyse problems of number theory* – that is, among others, distribution of primes, perfect numbers, Goldbach conjecture, etc.? And if this *is* possible then *how* to do it?

Indeed, such an approach has been suggested already roughly one hundred years ago by Hilbert and Poya, who conjectured that it may be possible to prove the Riemann hypothesis about the distribution of the non – trivial zeroes of the Riemann function $\zeta(s)$ by construction a Hermitian operator (we can speak about this as a Hamiltonian) whose eigenvalues in an appropriate sense correspond to those of $\zeta(s)$ (see, e. g. [1] for details and further references). Interesting review of what has been done in this direction can be found in [2]. In addition, recently we have witnessed a lot of original but not yet quite successful attempts to use closely related approach (to drop the self–adjointness requirement and search for the corresponding spectral properties of non-Hermitian "generalised annihilation operators"), see e. g. [3 - 6] and references cited therein.

Of course, to realise such a strategy, one needs first of all to get knowledge about possible spectral properties of different Hamiltonians with respect to the integers and other numbers of interest for the number theory. Quite surprisingly, but despite a huge literature dedicated to the spectral characteristics of Hamiltonians (inverse Sturm – Liouville problem, see e. g. [7]), it seems that such properties have practically never been viewed from this angle. The purpose of the current study is to analyse one relevant case: the construction of a Hamiltonian whose spectrum coincides with the set of primes is undertaken and results are discussed.

## 2. Power – like spectra

Let us consider the simplest one - dimensional Schrödinger equation $(-\frac{d^2}{dx^2} + V(x))\psi = \lambda\psi$ and ask for what spectra of $\lambda$ we can foresee existence of an appropriate potential $V(x)$. That is, we have a non degenerated and unbound from above spectrum $\lambda_1 < \lambda_2 < \ldots \lambda_i < \ldots$ and are interested what potential, if any, can give us this spectrum. Similar problems (inverse Sturm – Liouville problems on the whole line) have been extensively investigated in mathematics, and a number of conditions which should be imposed on the spectral matrix to ensure the existence of Sturm – Liouville operator has been found [7]. At the same time, the full solution of this problem is still unknown, and the mathematical results obtained seems can not be directly applied to our case. By this reason for us it seems natural (from physical reasoning) to start from the quasiclassical WKB approximation. It is clear that high – lying energy levels of the Hamiltonian at question (large $\lambda_i$) should be quite satisfactorily described by WKB approximation.

This approximation gives the following result concerning the number of energy levels smaller than a given value $E$ (which we will name a counting function):

$$N(E) = \frac{2}{\pi} \int_0^{x_{max}} \sqrt{2(E - V(x))}\,dx \qquad (1a)$$

Here and below we always suppose a symmetrical potential $V(x) = V(-x)$ and $x_{max}$ is such that $V(x_{max})=E$. Of course, numerical coefficients 2 and $\pi$ appearing in (1a) are not principal for our problem, so to simplify all the numerical coefficients below instead of this equation we will consider an equation

$$N(E) = 2\pi \int_0^{x_{max}} \sqrt{E - V(x)}\, dx \qquad (1)$$

Physically, it can be said that the unit system where $\dfrac{2m}{\hbar^2 \pi^4} = 1$ is used.

By differentiating (1) with respect to $E$ and changing the integration from $dx$ to $dV$ one obtains the following integral equation to determine the function $f(V) = dx/dV$ (c.f. the paper of Wu and Sprung [8]):

$$n(E) = \pi \int_{V_0}^{E} \frac{f(V)}{\sqrt{E-V}}\, dV \qquad (2)$$

Here $n(E) = dN(E)/dE$, and we will call it the density of eigenvalues; $V(0) = V_0$. This equation is nothing else than the famous Abel's integral equation whose solution is well known:

$$f(E) = \frac{n(V_0)}{\sqrt{E-V_0}} + \int_{V_0}^{E} \frac{n'(V)}{\sqrt{E-V}}\, dV \qquad (3).$$

Here $n'$ is a derivative of the density of eigenvalues: $n'(E) = dn(E)/dE$. An additive character of Eqs. (2, 3) is worthwhile to be noted: if $N(E) = N_1(E) + N_2(E)$ then $f(E) = f_1(E) + f_2(E)$ provided the same $V_0$ has been chosen when calculating both $f_1(E)$, $f_2(E)$.

First, it is natural to test (3) for the case of a Harmonic oscillator spectra $N(E)=aE+b$, thus $n(E)=a$, $n'(E)=0$ ($a>0$). Such a testing easily gives a desirable result $f(E) = dx/dE = a/\sqrt{E-V_0}$ and hence $E(x)=x^2/4a^2 + V_0$. This potential gives the eigenvalue distribution in question, moreover, for the case $a=1$ and $V_0=-1/2$ it gives even the exact correspondence $\lambda_i=i$. Of course, such an exact coincidence of spectra is accidental and can not be anticipated for the general case.

Second natural example is to search for the potential which gives a quadratic distribution of the eigenvalues of the type $\lambda_i$ = 1, 4, 9, 16, … Its "naturalness" is again motivated by quantum mechanics, from which it is well known that such a distribution is characteristic for the impenetrable square wall potential *V(x)=0* for -*l<x<l* and *V(x)=+∞* for *x≥l*, *x≤-l*, where one has a spectrum $\lambda_i = \pi^2(i+1/2)^2$. Setting $N(E) = 2\sqrt{E}$, $n(E) = 1/\sqrt{E}$, $n'(E) = -1/(2E^{3/2})$ we have:

$$f(V) = \frac{1}{\sqrt{V_0}\sqrt{E-V_0}} - \frac{1}{2}\int_{V_0}^{E} \frac{1}{V^{3/2}\sqrt{E-V}} dV \qquad (4).$$

Using variable change

$$E\text{-}V=t^2 \qquad (5)$$

(note, that this change eliminate any divergences from the integral), we after some trivial algebra obtain $f(E) = dx/dE = \dfrac{\sqrt{V_0}}{E\sqrt{E-V_0}}$. Its integration gives

$$E(x) = V_0(\tan^2(x/2)+1) \qquad (6).$$

Obviously, in some sense the potential (6) has a "limit" of the form of impenetrable square wall considered above when $V_0 \to 0$, $l = \pi$. Of course, for actual levels of the potential (6) one does not obtain exact squares of integers 1, 4, 9… It seems possible, however, to "fine tune" this potential to achieve the required correspondence with an arbitrary high accuracy.

It seems natural and is, of course, known that more "rarefied" distribution of the eigenvalues than the squares $\lambda_i$ = 1, 4, 9, 16, … given above can not be achieved for simple one – dimensional Schrödinger equation $(-\dfrac{d^2}{dx^2}+V(x))\psi = \lambda\psi$. (What potential can be more "effective" than the impenetrable square wall?). In WKB approximation this can be shown as follows. Let us start from the power – like distribution of eigenvalues of the type

$$N(E) = (1/2-\alpha)^{-1} E^{1/2-\alpha} \qquad (7)$$

where $0 < \alpha < \frac{1}{2}$. From (7) we have $n(E) = E^{-(1/2+\alpha)}$, $n'(E) = -(1/2+\alpha)E^{-(3/2+\alpha)}$. Repeating the same steps as above we obtain the following function $f(E)$:

$$f(E) = \frac{1}{V_0^{1/2+\alpha}\sqrt{E-V_0}} - (1+2\alpha)\int_0^{\sqrt{E-V_0}} \frac{dt}{(E-t^2)^{3/2+\alpha}} \qquad (8).$$

Our aim is to show that this function $f(E)$, which is equal to $dx/dE$, for any given $V_0$ starts to be negative when $E$ is large enough. This is, of course, unacceptable because then the potential $E(x)$ becomes a non single valued function of $x$ for sufficiently large $x$. Using the substitution $\cos\varphi = t/\sqrt{E}$, integral in (8) takes the form $-\frac{(1+2\alpha)}{E^{1+\alpha}} \int_{\arccos\sqrt{1-V_0/E}}^{\pi/2} \frac{d\varphi}{\sin^{2+2\alpha}\varphi}$ and, applying example #2.643 from [9]

$$\int \frac{d\varphi}{\sin^{2+2\alpha}\varphi} = -\frac{\cos\varphi}{(1+2\alpha)\sin^{1+2\alpha}\varphi} + \frac{2\alpha}{1+2\alpha}\int\frac{d\varphi}{\sin^{2\alpha}\varphi},$$ it can be finally reduced to the following expression:

$$f(E) = \frac{V_0^{1/2-\alpha}}{E\sqrt{E-V_0}} - \frac{2\alpha}{E^{1+\alpha}} \int_{\arccos\sqrt{1-V_0/E}}^{\pi/2} \frac{d\varphi}{\sin^{2\alpha}\varphi} \qquad (9).$$

Multiplying both sides of (9) by the definitely positive value of $E^{1+\alpha}$ we have

$$E^{1+\alpha} f(E) = \frac{V_0^{1/2-\alpha} E^{\alpha}}{\sqrt{E-V_0}} - 2\alpha \int_{\arccos\sqrt{1-V_0/E}}^{\pi/2} \frac{d\varphi}{\sin^{2\alpha}\varphi} \qquad (10)$$

from which the required property follows immediately: while the positive term in the right hand side of (10) for any $V_0$ tends to zero when $E \to \infty$ (remind that $0 < \alpha < \frac{1}{2}$), the negative term tends to a finite value of $-2\alpha \int_0^{\pi/2} \frac{d\varphi}{\sin^{2\alpha}\varphi}$ (this integral is, of course, a principal value in the sense of Cauchy).

Hence to find a Hamiltonian whose spectrum is a power – like and has a counting function growing more slowly than $N(E)=aE^{1/2}$ is impossible. (Of course, this is shown only for WCB approximation, but it seems obvious that if this is impossible for a quasi classic case this is impossible for the exact solution as well. Inverse implication is, of course, not at all evident). At the same time we can construct a Hamiltonian whose spectrum corresponds to an arbitrary large but finite set of numbers distributed as $n^l$, where $l > 2$: from (10) it is clear that if $V_0$ is taken to be large enough, the range of $E$ where r.h.s. of (10) is positive is also large. For example, for $\alpha=1/6$, what corresponds to the spectra $\lambda_i \propto i^3$, from (10) for large $V_0$ we need $E < V_o q^{-3} \approx 2.944 V_0$, here $q = \frac{1}{3}\int_0^{\pi/2}\frac{d\varphi}{\sin^{1/3}\varphi} \approx 0.698$. Evidently, an arbitrary large number of cubes $i^3$ can be found between $V_0$ and $2.944V_0$ as far as $V_0$ is large enough.

Correspondingly, no one potential can give a "multiplicative" eigenvalue spectrum of the type $\lambda_i = 1, p, p^2, ..., p^i, ...(p > 1)$. It can be easily illustrated directly: for such a spectrum we have $N(E) = \log_p(E)$, hence $n(E) = 1/(E\ln(p))$ and $n'(E) = -1/(E^2 \ln(p))$. For this eigenvalue distribution the determination of the function $f(E)$ again reduces to the standard integrals. The formal answer is:

$$\ln p \cdot f(E)E^{3/2} = \frac{1}{\sqrt{1-V_0/E}} - \frac{1}{2}\ln\left(\frac{(\sqrt{E}+\sqrt{E-V_0})^2}{V_0}\right) \quad (11),$$

from which the negativity of the resulting expression for any $V_0$ and large enough $E$ is evident.

From the same consideration it is clear that in WKB approximation any power – like eigenvalue distribution, which is denser than the square root of $E$, can be generated by an appropriate Hamiltonian. (Which, of course, seems self evident from the quantum mechanics). Starting from $N(E) = (1/2+\beta)^{-1}E^{1/2+\beta}$, where $\beta > 0$, one simply repeats one-to-one the same steps to obtain

$$f(E) = \frac{V_0^{1/2+\beta}}{E\sqrt{E-V_0}} + \frac{2\beta}{E^{1-\beta}} \int_{\arccos\sqrt{1-V_0/E}}^{\pi/2} \sin^{2\beta}\varphi \, d\varphi \qquad (12).$$

Whence *f(E)* is evidently always positive so nothing prevents the construction of a monotone function *x(E)* and its inverse:

$$x(E) = 2V_0^{\beta} \arctan\sqrt{\frac{E-V_0}{V_0}} + 2\beta \int_{V_0}^{E} \frac{dE}{E^{1-\beta}} \int_{\arccos\sqrt{1-V_0/E}}^{\pi/2} \sin^{2\beta}\varphi \, d\varphi \qquad (13).$$

## 3. What Hamiltonian can generate the primes: quasiclassical approach.

After elucidating the problem what power – like spectra can and what can not be generated in WKB approximation by a simple one – dimensional Hamiltonian, it is a time to consider a few examples. First, and in possible connection with the Riemann hypothesis, one can pose the problem of construction a Hamiltonian whose zeroes correspond to those of Riemann function $\zeta(s)$ (here speaking about zeroes of $\zeta(s)$ we mean the imaginary part $\tau$ of non trivial zeroes having (presumably) the form ½ $+i\tau$ [1]). According to the Riemann – van Mangoldt formula [1], distribution of such zeroes in the leading order is given by a function

$$N(E) = \frac{E}{2\pi} \ln\frac{E}{2\pi e} + O(\ln E) \qquad (14)$$

hence by a function which definitely grows rapidly enough to ensure the existence of the corresponding Hamiltonian.

This problem has been solved in [8]. For completeness, here we present function *x(E)* derived in [8] for the eigenvalue distribution (14); note that somewhat other than ours unit system, where $\frac{2m}{\hbar^2 \pi^2} = 1$, is used:

$$x = \frac{1}{\pi}\left(\sqrt{E-V_0}\ln\frac{V_0}{2\pi e^2} + \sqrt{E}\ln\frac{(\sqrt{E}+\sqrt{E-V_0})^2}{V_0}\right) \qquad (15)$$

This potential has been "fine tuned" by Wu and Sprung to give the best coincidence with the first 500 zeroes of $\zeta(s)$, and a fractal structure of the resulting potential with the dimension $d = 1.5$ was claimed. (Numerically it was shown earlier that the fluctuating part of the Riemann $\zeta(s)$ zeroes to obey Gaussian unitary ensemble statistics [10]).

What is very interesting and might look surprising (and for me it *does* look as such), nothing prevents search of a potential (Hamiltonian), whose spectrum $\lambda_i$ would generate the primes, that is for any $i$ $\lambda_i = p_i$ where $p_i$ is the *i*-th prime: $p_i = 2, 3, 5, 7,...$ As known, prime distribution $N(E)$ (designating in number theory as $\pi(E)$) in the leading order is given by integral logarithm $\pi(E) = li(E) = \int^E \frac{dt}{\ln t}$ [1], that is by the function which grows more rapidly than the square root $\sqrt{E}$.

Let us now search for such a Hamiltonian in quasiclassical approximation. From $\pi(E)$ we derive $n(E) = 1/\ln E$, $n'(E) = -1/(E \ln^2 E)$ and, proceeding as above, obtain the following equation for the determination of $f(E)$:

$$f(E) = \frac{1}{\ln V_0 \cdot \sqrt{E - V_0}} - \int_{V_0}^{E} \frac{dV}{V \ln^2 V \cdot \sqrt{E - V}} \qquad (16)$$

To avoid unnecessary complications we will suppose $V_0 > 1$. Observing that

$$\frac{d}{dV}\left(\frac{\ln E / \ln V - 1}{\sqrt{E - V} \ln E}\right) = \frac{\ln E / \ln V - 1}{2(E - V)^{3/2} \ln E} - \frac{1}{V \ln^2 V \sqrt{E - V}} \qquad (17),$$

(16) can be rewritten as

$$f(E) = \frac{1}{\ln V_0 \cdot \sqrt{E - V_0}} + \frac{\ln E / \ln V - 1}{\ln E \sqrt{E - V}}\Big|_{V_0}^{E} - \int_{V_0}^{E} \frac{\ln E / \ln V - 1}{2(E - V)^{3/2} \ln E} dV \qquad (18).$$

This operation is legitimate because $\lim_{V \to E} \frac{\ln E / \ln V - 1}{\ln E \sqrt{E - V}} = 0$. From (18)

$f(E) \ln E = \frac{1}{\sqrt{E - V_0}} - \int_{V_0}^{E} \frac{\ln E / \ln V - 1}{2(E - V)^{3/2}} dV$ follows trivially, and using variable change $\xi = V / E$

one has

$$f(E) \ln E \sqrt{E} = (1 - V_0 / E)^{-1/2} + \frac{1}{2} \int_{V_0/E}^{1} \frac{\ln \xi}{\ln(\xi E)} \cdot \frac{d\xi}{(1 - \xi)^{3/2}} \quad (19).$$

The easiest way to analyse the integral appearing in (19) is to note that the function $1/\ln(\xi E)$ in the integration area is a simple monotone function without any peculiarities and decreasing from $1/\ln V_0$ to $1/\ln E$. Without this function the integral is quite standard and well – behaving at both relevant limits $\xi = 0, 1$:

$$\frac{1}{2} \int \frac{\ln \xi}{(1 - \xi)^{3/2}} d\xi = \left( \frac{1}{\sqrt{1 - \xi}} - 1 \right) \ln \xi + 2 \ln(1 + \sqrt{1 - \xi}) \quad (20).$$

Hence $f(E) \ln E \sqrt{E} > (1 - V_0 / E)^{-1/2} - 2 \ln 2 / \ln V_0$ from which the positivity of $f(E)$ for any $E$ provided $V_0 > 4$ follows.

Let us define function $F(V_0, E)$ as a right-hand side of (19). For $E \to V_0$ this function is dominated by $1/\sqrt{1 - E/V_0}$ and thus is a divergent one. For large $x$ it logarithmically slowly, as $1 - C/\ln E$ where $C = 2\ln 2 \approx 1.386$, tends to unity. Numerical analysis shows that this function is everywhere positive for any $V_0 > 1$, not only for $V_0 > 4$ as it was demonstrated above. It has a minimum somewhere not so far from $V_0$ (for example, for $V_0 = 1.5$ the minimum is equal to $\approx 0.3$ and is attained at $E \approx 6$). Function $F(V_0, E)/(\ln E \sqrt{E})$ is simpler than $F(V_0, E)$ and monotonically decreases in all its area of definition $E > V_0$. An inverse square root – type divergence of this function at $E \to V_0$ does not prevent its integration which gives a monotone function $x(E)$:

$$x = \int_{V_0}^{E} \frac{F(V_0, E)}{\sqrt{E} \ln E} dE \tag{21}.$$

For (very) large $x$ we have:

$$x \approx A_1 + \int_{M}^{E} \frac{1}{\sqrt{E} \ln E} dE = A_2 + li(\sqrt{E}) \tag{22},$$

where $M$, $A_1$, $A_2$ are appropriate positive constants. Function $V_p(x)$, that is an inverse of $x(E)$ obtained by integration of (21) for $V_0 = 1.5$ is presented in Fig. 1.

Above we have used only the leading order of the prime counting function to demonstrate that the corresponding prime – generating potential exists. This result can be without serious difficulties generalised for some essentially more complicated "prime counting functions", for example for the Riemann $R$-function $\pi(E) = li(E) - \frac{1}{2} li(E^{1/2}) - \frac{1}{3} li(E^{1/3}) - ...$, which is known to describe better the first (many milliards of) primes [15]. We will not do this here simply not to repeat essentially the same as above. Heuristically, the possibility to use this counting function is due to the following: all terms of the type $\frac{1}{k} li(E^{1/k})$, where $k$ is an integer larger or equal to 2, grow less rapidly that the square root $E^{1/2}$. Hence, it is natural that, when applying Eq. (3) to a counting function of the type $N(E) = li(E^{1/k})$, one obtains the "impossible" case: for any $V_0$ and large enough $E$, $f(E)$ is negative. Correspondingly, because terms of such type have a minus sign in the Riemann $R$-function, the situation will be only better than that for the function $li(E)$: $f(E)$ will be even "more positive". Quite similarly, it can be shown that for some other counting functions as, for example $li(E) - C\sqrt{E} \ln E$, $C > 0$, no one suitable Hamiltonian can be found.

Indeed, all these examples going beyond the integral logarithm as a prime counting function do not demonstrate a lot: distribution of primes is *not* given by the $R$-function or like. (Probably the "quasiclassical" testing of exact prime counting function given by Riemann – van Mangoldt formula could be interesting, but this formula, implying infinite sums over the zeroes of $\zeta(s)$ [1, 2], seems too complicated to attempt such a testing). So we could limit ourselves with the $li(E)$, from which the real prime counting function deviates to both sides: definitely as $\Omega_\pm(\frac{\sqrt{E}}{\ln E}\ln\ln\ln E)$ (Littlewood's result), no more than $O(E\exp(-C\ln^{3/5} E \ln\ln^{-1/5} E))$, $C > 0$, (the best estimation proven) or than $O(\sqrt{E}\ln E)$ provided the Riemann hypothesis holds true [1]. Recall that, as we have seen earlier for the example of the spectrum of cubes, deviations from a leading counting function described by an "impossible" distribution law can be tolerated at long intervals. (Let us also remind here still unproven but highly believed-to-be-true conjecture that there is at least one prime between any two squares of the neighbouring integers, $i^2$ and $(i+1)^2$. If this holds true, this fact could be regarded as an additional informal argument in favour of existence of the Hamiltonian in question). To conclude this Section, we can only say that from the quasiclassical consideration it follows that to try to search numerically for a Hamiltonian whose spectrum coincides with the set of primes does make a sense.

## 4. What Hamiltonian can generate the primes: numerical results.

To construct the required Hamiltonian I used the method proposed by Ramani *et al.* [11, 12]. They showed that it is often possible to construct a Hamiltonian having the spectrum $\lambda_1 < \lambda_2 <\ldots <\lambda_N=0$ using the following procedure. One starts from some Schrödinger equation $-\Psi'' +V(x)\Psi = \varepsilon\Psi$ having the deepest eigenvalue $\varepsilon_0$. (Of course, it is natural to set simply $V(x) = 0$ and hence $\varepsilon_0 = 0$). If we chose now some $\varepsilon_1 < \varepsilon_0$ than one can find numerically such a solution $\Psi$ of this equation, which diverges exponentially when $x \to \pm\infty$ and has no node in between. Obviously, the function $\Phi = 1/\Psi$ decreases exponentially at both infinities, has no divergences in between and can be normalised. It is not difficult to show that this function is a solution of Schrödinger equation with the following potential $W$:

$$W(x) = 2\varepsilon + 2f^2 - V(x) \qquad (23).$$

Here $f$ is a logarithmic derivative of $\Psi$: $f = -\Psi'/\Psi$. Ramani *et al*. noted that (23) is precisely the so called dressing transformation, for which it has been shown that modifying a potential $V$ in just such a way one obtains the potential whose spectrum contains both $\varepsilon_0$, $\varepsilon_1$ (whence "dressing", see e. g. [13]). Further, starting with the $W$ as a new potential one should put new eigenvalue $\varepsilon_2 < \varepsilon_1$ to obtain the next potential $W_1$ with the spectrum $\varepsilon_2$, $\varepsilon_1$, $\varepsilon_0$, and so forth.

Following this procedure, to construct a Hamiltonian whose spectrum coincides with the set of the first 500 (or other number) of primes we first generate the set of primes $p_1 = 2, \ldots, p_{500} = 3571$. Then we subtract $p_{500}$ from all members of this set to obtain the set $p_1 - p_{500}, p_2 - p_{500}, \ldots, 0$. Finally, we reverse its order obtaining $0, p_{499} - p_{500}, \ldots, p_1 - p_{500}$, and apply the procedure described above to this new set: $\varepsilon_0 = 0$, $\varepsilon_1 = p_{499} - p_{500}$, $\varepsilon_2 = p_{498} - p_{500}$, etc. This results in a potential whose 500 eigenvalues, after adding $p_{500}$, coincides with the first 500 primes. It is evident that such a numerical procedure is effective and can be easily implemented at any modern PC for rather long sets of eigenvalues.

In Fig. 2 I present the potentials $V$ obtained for the first 500 and 1000 primes. No difficulties occur during the calculations on a standard Pentium PC using MATLAB packet (of course, some attention should be paid on the calculation precision (Runge - Kutta method step size) to ensure the convergence of calculations; calculations for 1000 primes took overnight but no serious attempts to write the most effective algorithm have been undertaken). No signs of "catastrophic divergences", similar to those reported in [11] when the construction of a potential generating an "impossible" spectrum $\lambda_i = i^3$ has been attempted, were noticed. Fig. 2 clearly reveals the "fractal" nature of the Hamiltonian generating the primes, in the sense that the period of the potential quasioscillations (mean distance between the neighbouring extrema or like) decreases when the number of primes to be generated increases. However, I did not attempt to determine the fractal dimension of this dependence (using, e. g., counting box method) because such a procedure is known to converge very slowly (cf. [8, 11, 12, 14]) and, more importantly, it is not clear what type of information can be extracted from this data. Comparison of the prime – generated potential calculated numerically with the "quasiclassical" potential $V_p(x)$, similar to that presented in Fig. 1, reveals that potential $V_p(x)$ quite a satisfactorily describes the general trend of exact "fractal" potential. Analysis also reveals that function $V$ has especially complex character in certain regions (for example, near $x$ = 8 - 9), which in all probability is due to the "most irregular" distribution of primes for corresponding regions of $E$ (for the case given above – around 500).

Briefly, numerical data demonstrates that the Hamiltonian whose spectrum coincides *exactly* with the set of primes can be constructed for large sets of primes.

## 5. Discussion and conclusion

Of course, consideration given above is by no means a proof that there exists a Hamiltonian which admits solutions from $L^2(-\infty, +\infty)$ iff $\lambda_i = p_i = 2, 3, 5, \ldots$ Nevertheless, we hope that we gave some evidences that existence of such a Hamiltonian could be regarded as a highly probable. I do not also really know how such a Hamiltonian, provided it does exist and this is proven, could be used to study the prime – related problems of number theory. Nevertheless, I would like to conclude the paper with the following qualitative illustration which, I believe, might stimulate further research in this direction.

Let us consider the following problem. We have a Hamiltonian whose spectrum is described in the leading order by the integral logarithm, that is *average* density of its eigenvalues decreases as *1/lnE*. At the same time, let us require its spectrum to contain additionally an infinite number of *twins* – that is, neighbouring eigenvalues which differ from each other by unity. The question is *how many twins can be tolerated*, in other words how many twin eigenvalues can be generated by a Hamiltonian whose spectrum counting function is still asymptotically given by *li(E)*?

Obviously, this question has connections with the famous twin primes problem [1, 15]: how many there are primes which differs from each other by 2, how they are distributed, etc. Of course, here we consider not the "fractal" and very complicated exact prime - generating potential *V(x)* (see Section 4) but treat *V(x)* as a simple monotone function. Again, we hope that if something is impossible in the frame of quasiclassical approach, it should be impossible also for the full and strict treatment.

Now (1) can be regarded as a quasiclassical quantization condition, from which from time to time we obtain neighbouring levels $E_N$ and $E_{N+1}=E_N+1$:

$$N = 2\pi \int_0^{x_{max1}} \sqrt{E_N - V(x)}\,dx \qquad (24a)$$

$$N+1 = 2\pi \int_0^{x_{max2}} \sqrt{E_N + 1 - V(x)}\,dx \qquad (24b).$$

Asymptotical dependence of *V(x)* on *x* as inverse of the integral logarithm of $V^{1/2}$, which has been shown in Section 3 for the counting function in hand, determines that the difference between $x_{max1}$ and $x_{max2}$ can not be too small. By the differentiation of (24) with respect to *E* and changing the variable from *x* to *V* we have, similarly as above to derive Eq. (2):

$$\frac{dN}{dE} = \pi \int_{V_0}^{E} \frac{1}{\sqrt{E-V}} \frac{dx}{dV} dV \cong \pi \int_{V_0}^{E} \frac{dV}{\sqrt{E-V}\sqrt{V} \ln V} \qquad (25).$$

Here we use for $x(V) = li(V^{1/2})$ the relation $dx/dV = 1/(\sqrt{V} \ln V)$. Using again the variable change $\xi = V/E$, (25) can be rewritten as

$$\frac{dN}{dE} = \pi \int_{V_0/E}^{1} \frac{d\xi}{\sqrt{1-\xi}\sqrt{\xi} \ln(\xi E)} \qquad (26).$$

Because the integral value $\int_0^1 (1-\xi)^{-1/2} \xi^{-1/2} d\xi = \pi$ is finite, from (26) it follows that the value of the derivative $dN/dE$ for large $E$ tends to zero asymptotically as $\pi^2/\ln E$. This means that for large $E$ only a very small increment of $N$ can be produced by the function $\sqrt{E_N + 1 - V(x)}$ integrated from 0 to $x_{max1}$ and, correspondingly, $\pi \int_{x_{max1}}^{x_{max2}} \sqrt{E_N + 1 - V(x)} dx \cong 1$. Because in the integration region the value of expression under the integral sign does not exceed 1, for large $E$ one has $x_{max2} - x_{max1} \geq 1/\pi$ whatever local changes of $V(x)$ have been done in the region between $x_{max1}$ and $x_{max2}$. (Indeed, this is valid for any monotone potential $V(x)$ growing faster than $Cx^2$). In other words, to accommodate the number $N_{tw}(E)$ of twins less than or equal to $E$, we need $x \geq N_{tw}(E)/\pi$. From this it follows, because for our spectral counting function we suppose $x \cong li(\sqrt{E})$, that the twin eigenvalues counting function $N_{tw}(E)$ should be at most $O(li(\sqrt{E}))$ and hence the twin density $n_{tw}(E)$ at most $O(1/(\sqrt{E} \ln E))$.

Note, that these values are smaller than those which are usually conjectured for twin primes, for which density $n_{tw} = C/\ln^2 E$ is supposed (this is the Hardy - Littlewood conjecture supported, among others, by simple "probability" argument, cf. [15]). The number of twin primes is still infinite, but not only the sum of reciprocal twin primes $\sum 1/p$ should be finite (what is proven by Brun theorem [16]) but all the sums $\sum 1/p^{\alpha}$, ($\alpha$>1/2) should also be finite for the twin primes (twin eigenvalues) distribution obtained, $N_{tw}(E) = O(li(\sqrt{E}))$.

Thus from existence of the prime – generating Hamiltonian some new and unexpected properties of twin primes can be at least conjectured, which is, I believe, already something interesting. (Of course, from somewhat different point of view, above consideration of the twin eigenvalues problem could be considered simply as a heuristical argument *against* existence of a prime – generating Hamiltonian. But, given that nothing is really known about the twin primes (it is not even proven that their number is infinite [1, 15]), its importance is not so large). To conclude, I would like again to express hope that the consideration undertaken in the paper is interesting for some readers and could stimulate somebody for further work in this direction.

**Figures**

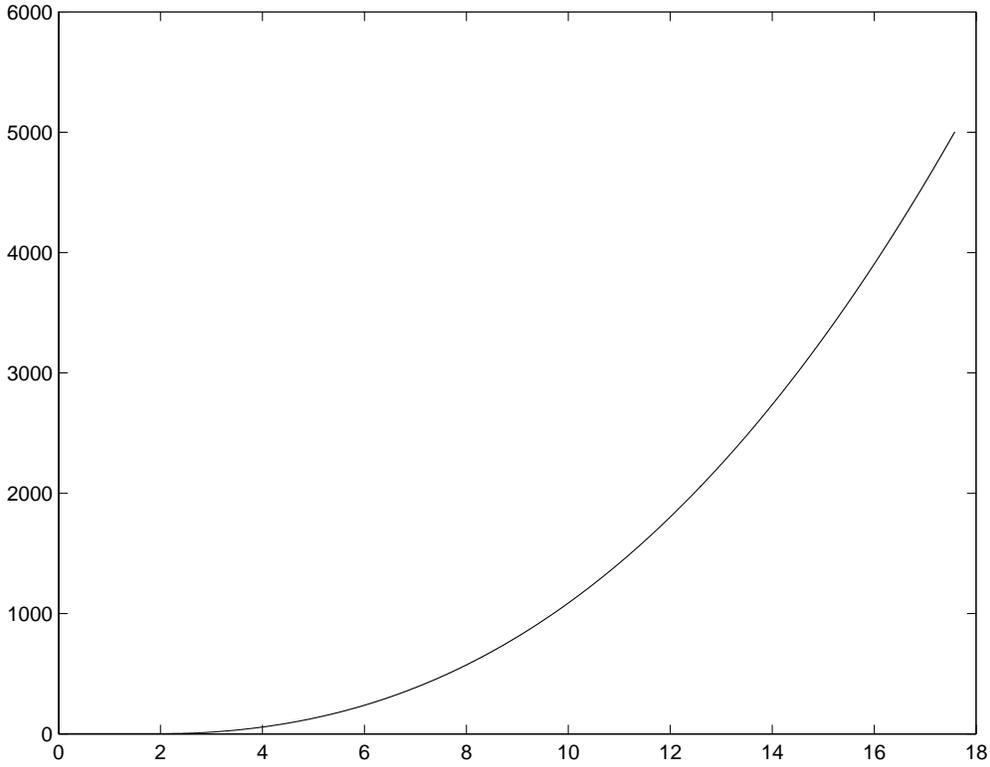

Fig. 1. Potential $V_p(x)$ whose spectrum has a counting function $li(E)$. Here shown for $V_0 = 1.5$.

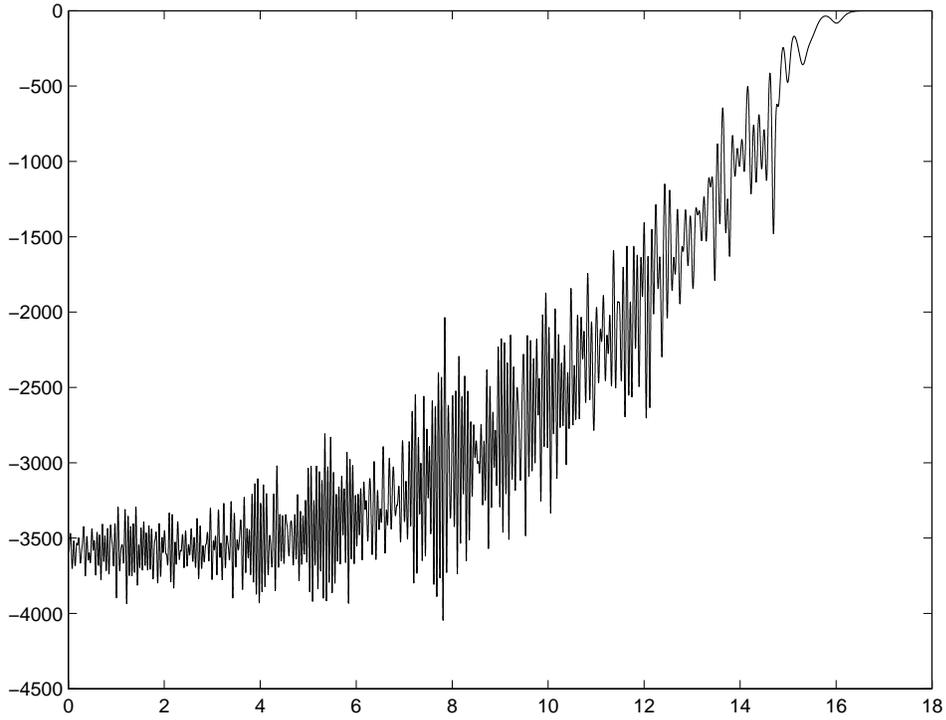

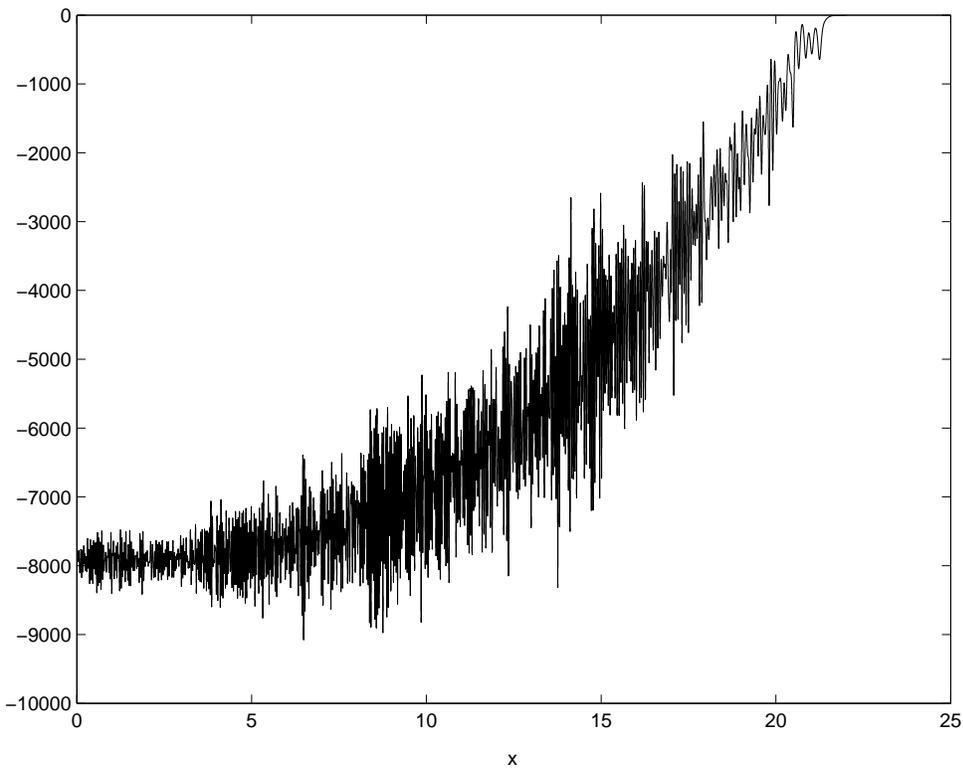

Fig. 2. Potentials *V(x)*, whose spectra coincide with the first 500 (up) and 1000 (down) primes.